\documentclass[aps,pra,amsmath,amssymb,twocolumn,superscriptaddress,notitlepage,reprint]{revtex4-1}
\pdfoutput=1
\usepackage{graphicx}
\usepackage{color}
\definecolor{blue}{RGB}{25,51,170}
\definecolor{green}{RGB}{51, 170, 42}
\definecolor{red}{RGB}{238,25,25}
\usepackage[colorlinks,citecolor=blue,linkcolor=red,urlcolor=green]{hyperref}
\hypersetup{
  pdfkeywords={},
  pdftitle={Theory of a Quantum Scanning Microscope for Cold Atoms},
  pdfauthor={D. Yang, C. Laflamme, D.V. Vasilyev, M. A. Baranov, P. Zoller}}

\makeatletter

\newcommand{\ket}[1]{\left|#1\right\rangle}
\newcommand{\bra}[1]{\langle#1|}

\newcommand{\be}{\begin{equation}}
\newcommand{\ee}{\end{equation}}
\newcommand{\ba}{\begin{eqnarray}}
\newcommand{\ea}{\end{eqnarray}}

\makeatother

\begin{document}

\title{Theory of a Quantum Scanning Microscope for Cold Atoms}

\author{D. Yang}

\author{C. Laflamme}

\author{D. V. Vasilyev}

\author{M. A. Baranov}

\author{P. Zoller}

\address{Institute for Theoretical Physics, University of Innsbruck, A-6020
Innsbruck, Austria}

\address{Institute for Quantum Optics and Quantum Information of the Austrian
Academy of Sciences, A-6020 Innsbruck, Austria}

\date{\today}
\begin{abstract}
We propose and analyze a scanning microscope to monitor `live' the quantum dynamics
of cold atoms in a Cavity QED setup. The microscope measures the atomic
density with subwavelength resolution via dispersive couplings to
a cavity and homodyne detection within the framework of continuous
measurement theory. We analyze two modes of operation. First, for
a fixed focal point the microscope records the wave packet dynamics
of atoms with time resolution set by the cavity lifetime. Second,
a spatial scan of the microscope acts to map out the spatial density
of stationary quantum states. Remarkably, in the latter case, for a good cavity limit, the
microscope becomes an effective quantum non-demolition (QND) device, such
that the spatial distribution of motional eigenstates can be measured
back-action free in single scans, as an emergent QND measurement. 
\end{abstract}

     
\maketitle
Spatially resolved observation of individual atoms is a key ingredient
in exploring quantum many-body dynamics with ultracold atoms. This
is highlighted by the recent development of the quantum gas microscope~\cite{*[{For a review, see }] [{ and references therein.}] Kuhr2016}
where fluorescence measurements provide us with single shot images
of atoms in optical lattices. Fluorescence imaging is, however, an
inherently destructive quantum measurement, as it is based on multiple
resonant light scattering resulting in recoil heating (see, however,~\cite{Ashida2015}). In contrast,
quantum motion of cold atoms can also be observed in non-destructive, weak measurements, realizing the paradigm of \textit{continuous measurement} of a quantum system~\cite{Braginsky1992,wiseman2009quantum,gardiner2015quantum}.
Below we describe and analyze a quantum optical setup for a \textit{scanning
atomic microscope} employing dispersive interactions in a Cavity QED
(CQED) setup~\cite{*[{For recent reviews, see }] [{}] Mekhov2012,*Ritsch2013,*Northup2014,*Rempe2015}, where the goal is to achieve continuous
observation of the density of cold atoms~\cite{*[{Continuous observation of classical atomic motion in Cavity QED is demonstrated in~}] [{}] Hood1447, *Puppe2007,*Terraciano2009} with subwavelength resolution~\cite{*[{Strong measurement of quantum particles with sub-wavelength resolution is demonstrated, e.g. in~}] [{}] Maurer2010, *Blatt2010}.
We will be interested in operating modes, where we either map out spatial densities of energy eigenstates in \emph{single scans} as an \emph{emergent} QND measurement~\cite{comment1, *[{For QND measurements in quantum optics, see, e.g., }][{}] Gleyzes2007,*Johnson2010,*Volz2011,*Barontini2015,*Moller2017},
or we monitor at a fixed position, the time resolved response to `see'
quantum motion of atoms.

\begin{figure}[t!]
\centering{}\includegraphics[width=0.49\textwidth]{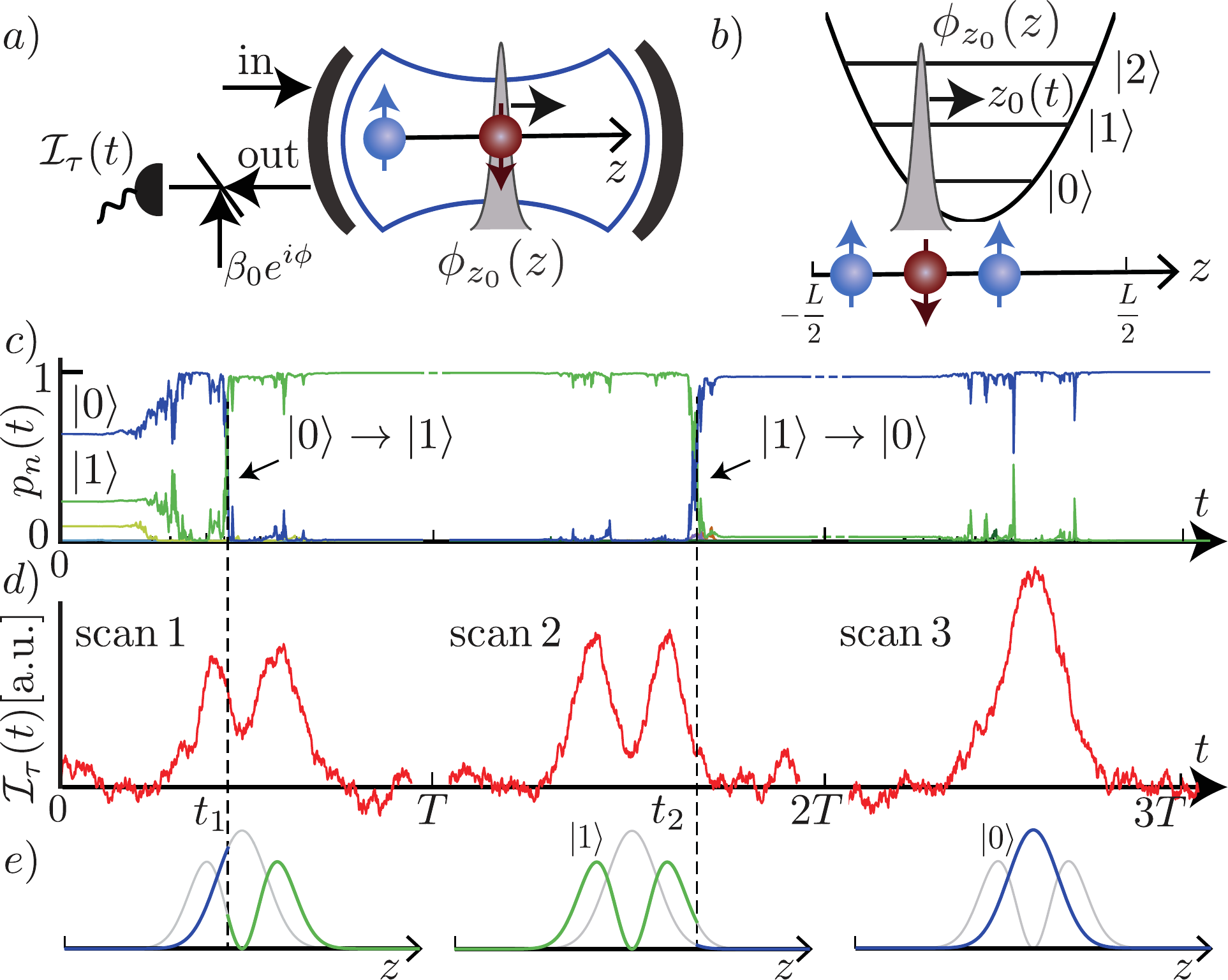} \caption{(a) Scanning Microscope as a CQED setup: The atom signals its presence in the focal region with subwavelength resolution as a spin flip, detected via a dispersive cavity coupling in homodyne measurement. (b) {\em Spatial scan} of the focal point $z_0=z_0(t)$ for an atom in a harmonic oscillator (HO). (c-e) Operation of the microscope in the good cavity limit as emergent QND measurement (see text). For an atom in a thermal state of the HO we {\em simulate  a single measurement run} involving three consecutive spatial scans:  (c) conditional trap populations $p_n(t)$ ($n=0,1,2$), and (d) homodyne current  ${\cal I}_{\tau}(t)$. QND measurement prepares the atom in a  trap state $\ket{n}$, and ${\cal I}_{\tau}(t)$ traces the corresponding density (e) in the subsequent scan. Times $t_1,t_2$ indicate quantum jumps between trap states (see text). 
}
\label{fig:figure1} 
\end{figure}

The operating principle of the microscope is illustrated as a CQED
setup in Fig.~\ref{fig:figure1}: We assume that an atom traversing
the focal region of the microscope signals its presence with an internal
spin flip, i.e.~the position, and thus motion of the atom, is correlated
with its internal spin degree of freedom. While subwavelength spatial
resolution can in principle be achieved by driving transition between
spin states in the presence of external fields generating energy shifts
with strong spatial gradients~\cite{Thomas1991,Weitenberg2011b}, this spatial resolution is typically accompanied with strong forces acting on the atom. Instead we will
describe below a setup with diminished disturbance, based on \textit{ position dependent `dark
state'} in a $\Lambda$-system~\cite{Gorshkov2008}, involving a
pair of longlived atomic ground state levels, representing the spin. We can detect this spin flip nondestructively
with a dispersive interaction, e.g.~as shift of a cavity mode of
an optical resonator. Thus the atom traversing the focal region of
the microscope, as defined by lasers generating the atomic dark state,
becomes visible as a phase shift of the laser light reflected from
the cavity. This phase shift is revealed in homodyne detection. Such CQED schemes are timely in view of both the recent progress with cold atoms in cavity and nano-photonic setups~\cite{Stute2012a,Wolke75,Hacker2016a,Leonard2017,Kollar2017,Thompson1202,Rauschenbeutel2013,Goban2014,Polzik2016}, and the growing interests in conditional dynamics of cold atoms under measurement~\cite{Steck2004,Lee2014,Wade2015,Mazzucchi2016,Ashida2017,Laflamme2017}.

Below we will develop a quantum optical model of continuous measurement~\cite{carmichael1993open,wiseman2009quantum,gardiner2015quantum}
of atomic density, via measurement of the homodyne current for the setup
described in Fig.~\ref{fig:figure1}. We adopt the language of the
Stochastic Master Equation (SME)
for the conditional density matrix $\rho_{c}(t)$ of the joint atom-cavity
system, which describes time evolution conditional to observation
of a given homodyne current trajectory, as `seen' in a single run
of an experiment, and including the backaction on the atom. This will
allow us to address to what extent the observed homodyne current in
a spatial scan provides a faithful measurement of atomic density,
and the expected signal-to-noise ratio (SNR). 

\emph{Quantum Optical Model} \textendash{} We consider a model system
of an atom moving in 1D along the $z$-axis, placed in a
driven optical cavity. To detect the atom at $z_{0}$ with resolution
$\sigma$ we introduce a spatially localized dispersive coupling of
the atom to a single cavity mode of the form 
\begin{equation}
\hat{H}_{{\rm coup}}=\phi_{z_{0}}(\hat{z})\,\hat{c}^{\dagger}\hat{c}.\label{eq:dispersive_coup}
\end{equation}
Here $\phi_{z_{0}}(z)$ defines a sharply peaked \emph{focusing function}
of support $\sigma$ around $z_{0}$, and $\hat{c}^{\dagger}\hat{c}$
is the photon number operator for the cavity mode with destruction
(creation) operators $\hat{c}\,(\hat{c}^{\dagger})$. An implementation
of $\phi_{z_{0}}(\hat{z})$ achieving optical subwavelength resolution
$\sigma\ll\lambda$ based on atomic dark states in a $\Lambda$-system
will be described below. We find it convenient to write $\phi_{z_{0}}(z)\equiv{\cal A}f_{z_{0}}(z)$
with $f_{z_{0}}(z)$ normalized, and $\mathcal{A}$ a constant with
the dimensions of energy. 

According to Eq.~\eqref{eq:dispersive_coup}, the presence of an
atom inside the focal region results in a shift of the cavity resonance.
This can be detected with homodyne measurement, where the output
field of the cavity is superimposed with a local oscillator with phase
$\phi$. The homodyne current can, for a single measurement trajectory,
be written as $I(t)=\sqrt{\kappa}\langle\hat{X}_{\phi}\rangle_{c}+\xi(t)$,
i.e.~follows the expectation value of the quadrature operator of
the intracavity field, $\hat{X}_{\phi}\equiv e^{i\phi}\hat{c}^{\dagger}+e^{-i\phi}\hat{c}$,
up to the (white) shot noise $\xi(t)$. Here $\kappa$ represents
the cavity damping rate, and $\langle\dots\rangle_{c}\equiv\mathrm{Tr}\{\ldots\,\rho_{c}(t)\}$
refers to an expectation value with respect to the conditional density
matrix of the joint atom-cavity system.

On a more formal level, we write for the evolution under homodyne
detection the It\^{o} stochastic differential equations for the homodyne
current 
\begin{align}
dX_{\phi}(t) & \equiv I(t)dt=\sqrt{\kappa}\langle\hat{X}_{\phi}\rangle_{c}dt+dW(t),\label{eq:homodyneCurrentDef}
\end{align}
with $dW(t)$ Wiener noise increments, and the SME for the conditional
density matrix 
\begin{align}
d\rho_{c} & \!=\!-\frac{i}{\hbar}\![\hat{H},\!\rho_{c}]dt\!+\!\kappa{\cal D}[\hat{c}]\rho_{c}dt\!+\!\sqrt{\kappa}{\cal H}[\hat{c}e^{-i\phi}]\rho_{c}dW\!(t).\label{eq:SME_general}
\end{align}
Eq.~(\ref{eq:homodyneCurrentDef}) identifies the homodyne current
as measurement of the quadrature component $dX_{\phi}(t)$ of the output field in a time step $[t,t+dt)$. The SME~\eqref{eq:SME_general}
contains the total Hamiltonian $\hat{H}=\hat{H}_{{\rm sys}}+\hat{H}_{{\rm c}}+\hat{H}_{{\rm coup}}$
with $\hat{H}_{{\rm sys}}=\hat{p}_{z}^{2}/2m+V(\hat{z})$ the Hamiltonian
of the atomic system in an external potential $V$, $\hat{H}_{{\rm c}}=i\hbar\sqrt{\kappa}\mathcal{E}(\hat{c}-\hat{c}^{\dagger})$
the Hamiltonian for the driven cavity in the rotating frame (we
assume resonant driving for simplicity), and ${\cal E}$ the coherent
amplitude of the cavity mode driving field. The last two terms in Eq.~\eqref{eq:SME_general} account
for the back-action of homodyne measurement. The Lindblad operator
$\mathcal{D}[\hat{c}]\rho\!\equiv\!\hat{c}\rho\hat{c}^{\dagger}-\frac{1}{2}\hat{c}^{\dagger}\hat{c}\rho-\frac{1}{2}\rho\hat{c}^{\dagger}\hat{c}$
describes the system decoherence (the cavity field damping) due to
the coupling to the outside electromagnetic modes, and the nonlinear
operator ${\cal H}[\hat{c}]\rho_{c}\equiv\hat{c}\rho_{c}-\langle\hat{c}\rangle_{c}\rho_{c}+\text{H.c.}$
updates the density matrix conditioned on the observation of the homodyne
photocurrent $I(t)$.

The relation between the homodyne current and the local atomic density is most transparent in the limit where the cavity response time $\tau_{c}=1/\kappa$ is much faster than other time scales including atomic motion $\hat{H}_{{\rm sys}}$, and the dispersive coupling $f_{z_{0}}$, i.e.~the bad cavity limit.
Adiabatic elimination of the cavity gives 
\begin{equation}
dX_{\phi}(t)\!\equiv\!I(t)dt\!=\!2\sqrt{\gamma}\langle f_{z_{0}}\!(\hat{z})\rangle_{c}\,dt\!+\!dW(t),\label{eq:homodyne_elimination}
\end{equation}
with the atomic conditional density matrix $\tilde{\rho}_{c}(t)$
obeying the SME 
\begin{align}
d\tilde{\rho}_{c} & =-\frac{i}{\hbar}[\hat{H}_{{\rm sys}},\tilde{\rho}_{c}]\,dt\!+\!\gamma{\cal D}[f_{z_{0}}(\hat{z})]\tilde{\rho}_{c}\,dt\nonumber \\
 & \quad+\sqrt{\gamma}{\cal H}[f_{z_{0}}(\hat{z})]\tilde{\rho}_{c}\,dW(t).\label{eq:SME_elimination}
\end{align}
Here $\gamma=[4\mathcal{A}\mathcal{E}/(\hbar\kappa)]^{2}$ is an effective measurement rate, and we have chosen $\phi=-\pi/2$ (see Appendix~\ref{app:adiabatic_elimination}). According to Eq.~(\ref{eq:homodyne_elimination})
the homodyne current $I(t)$ is a direct probe of the local atomic
density at $z_{0}$ with spatial resolution $\sigma$~\cite{*[{We contrast this to schemes where $I(t)$ reflects the atomic positon $\langle \hat{z}\rangle_c$ in, e.g. }][{}] Quadt1995,*1367-2630-12-4-043038}.
Eqs.~(\ref{eq:homodyne_elimination})
and (\ref{eq:SME_elimination}), or (\ref{eq:homodyneCurrentDef})
and (\ref{eq:SME_general}) in the general case, provide us with the
tools to study dynamics of the `microscope' in various modes of operation
(see below).

Instead of single trajectories, we can also consider ensemble averages
corresponding to repeated preparation and measurement cycles. We define
a density operator for the atom-cavity system $\rho(t)\!=\!\langle\rho_{c}(t)\rangle_{{\rm st}}$
as statistical average over the conditional density matrices, and an
averaged homodyne current $\langle I(t)\rangle_{{\rm st}}\!=\!\sqrt{\kappa}{\rm Tr}\{\hat{X}_{\phi}\rho(t)\}$.
This density operator obeys a master equation (ME), obtained
from the SME (\ref{eq:SME_general}) by averaging over trajectories.
Thus $\rho_{c}(t)\rightarrow\rho(t)$ in Eq.~\eqref{eq:SME_general}
with the stochastic term dropped according to the It\^{o} property $\langle\ldots\,dW(t)\rangle_{{\rm st}}=0$.
An analogous ME for the atom $\tilde{\rho}(t)=\langle\tilde{\rho}_{c}(t)\rangle_{\rm st}$
can be derived from the adiabatically eliminated SME (\ref{eq:SME_elimination}), see Appendix~\ref{app:adiabatic_elimination}.

\begin{figure}[t!]
\centering{} \includegraphics[width=0.46\textwidth]{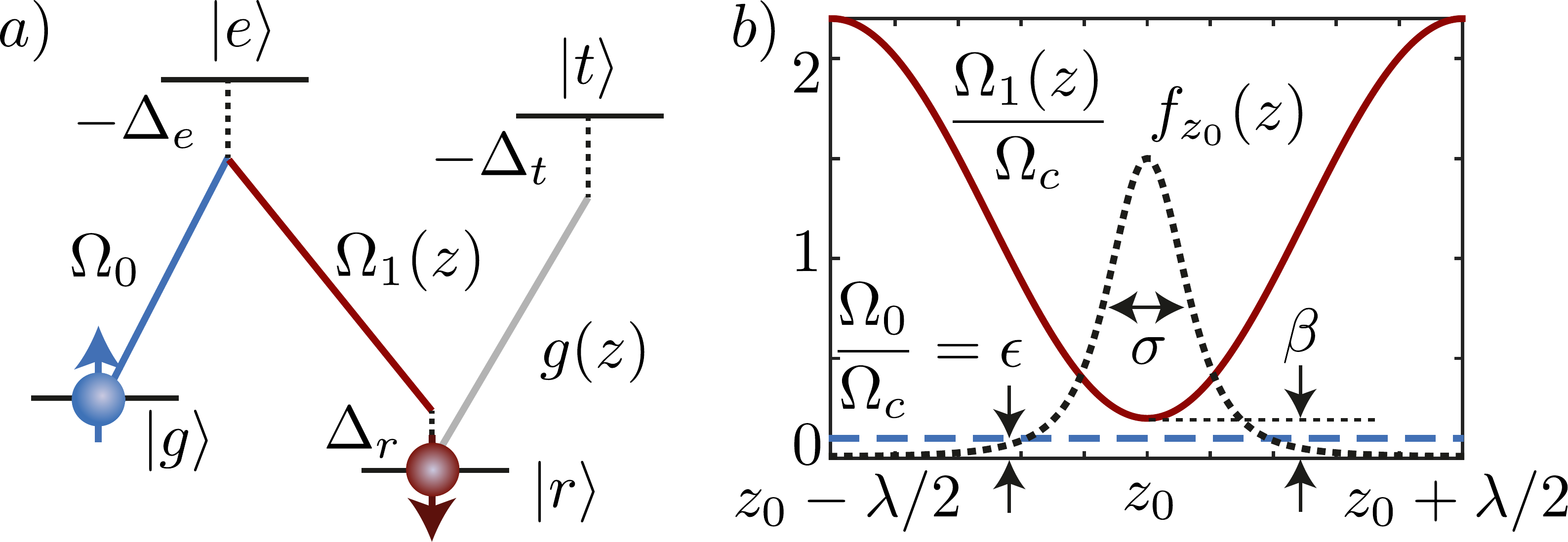} 
\caption{Implementing the focusing function $\phi_{z_{0}}(z)$. (a) The $\Lambda$-configuration $|g\rangle,|r\rangle,|e\rangle$ supporting a dark state
with a sub-wavelength spin structure associated with the ground states (see text), and dispersive cavity coupling on the transition $\ket{r}\rightarrow\ket{t}$. (b) The Rabi frequencies, $\Omega_{1}(z)=\Omega_{c}\{1+\beta+\sin[k(z-z_{0})]\}$ (solid line), $\Omega_{0}=\epsilon\Omega_{c}$ (dashed), and the (dimensionless) focusing function $f_{z_{0}}(z)$ (dotted) shown for $\epsilon=\beta/2=0.1$. For this configuration, the corresponding non-adiabatic potential \protect\cite{Lacki2006,Gorshkov2016} is strongly suppressed (see Appendix~\ref{app:FocusFunction}).}
\label{fig:fig2} 
\end{figure}

\emph{Implementation of the focusing function} $\phi_{z_{0}}(\hat{z})$
\textendash{} The atom-cavity coupling Eq.~\eqref{eq:dispersive_coup}
with subwavelength resolution can be achieved using the position-dependent
dark state of a $\Lambda$-system~\cite{Gorshkov2008,Lacki2006,Gorshkov2016}.
We consider the level scheme of Fig.~\ref{fig:fig2}a, where two
atomic ground (spin) states $\ket{g}$ and $\ket{r}$ are coupled
to the excited state $\ket{e}$ with Rabi frequencies $\Omega_{0}$
and $\Omega_{1}(z)$, respectively. This configuration supports a
dark state $\ket{D(z)}=\sin\theta(z)\ket{g}-\cos\theta(z)\ket{r}$
with $\tan\theta(z)=\Omega_{1}(z)/\Omega_{0}$, which via destructive interference is
decoupled from the dissipative excited state $\ket{e}$. We note that
in spatial regions $\Omega_{1}(z)\gg\Omega_{0}$ the atom will be
(dominantly) in state $\ket{g}$, while in regions $\Omega_{1}(z)\ll\Omega_{0}$
the atom will be in $\ket{r}$. This allows us to define via the spatial
dependence of $\Omega_{1}(z)$ regions with subwavelength resolution
$\left|z-z_{0}\right|\lesssim\sigma\ll\pi/k=\lambda/2$, characterized
by atoms in $\ket{r}$. Atoms in $\ket{r}$ can be dispersively coupled
to the cavity mode, resulting in a shift $g^{2}(z)/\Delta_{t}\hat{c}^{\dagger}\hat{c}$,
with $g(z)$ the cavity coupling much smaller than the detuning $\Delta_{t}$
(c.f. Fig.~\ref{fig:fig2}a). Thus atoms prepared in the dark state
experience a shift \eqref{eq:dispersive_coup} with 
\[
\phi_{z_{0}}(z)\!\equiv\!{\cal A}f_{z_{0}}(z)\!=\!\frac{\hbar g^{2}(z)}{\Delta_{t}}|\langle r|D(z)\rangle|^{2}\!=\!\frac{\hbar g^{2}(z)}{\Delta_{t}}\cos^{2}\theta(z).
\]
We illustrate this focusing function with subwavelength resolution
in Fig.~\ref{fig:fig2}b for a specific laser configuration.

\emph{Microscope Operation} \textendash{} The parameters characterizing
the microscope are the spatial resolution $\sigma\ll\lambda$, the
temporal resolution $\tau_{c}$ (given essentially by the cavity linewidth
$1/\kappa$) and the dispersive atom-cavity coupling controlling the
strength of the measurement. To be specific we illustrate below
the operation of the microscope as continuous observation of an atom
moving in a harmonic oscillator (HO) potential
with an oscillation frequency $\omega$ and vibrational eigenstates $\ket{n}$
($n=0,1,\ldots)$. The generic physical realization
includes a neutral atom in an optical trap (lattice),
or an ion in a Paul trap~\cite{gardiner2015quantum}, where we require a spatial
resolution better than the length scale set by the HO ground state
$\sigma\lesssim \ell_0=\sqrt{\hbar/m\omega}$ with $m$ the atomic
mass.

\begin{figure}[t]
\centering{} \includegraphics[width=0.49\textwidth]{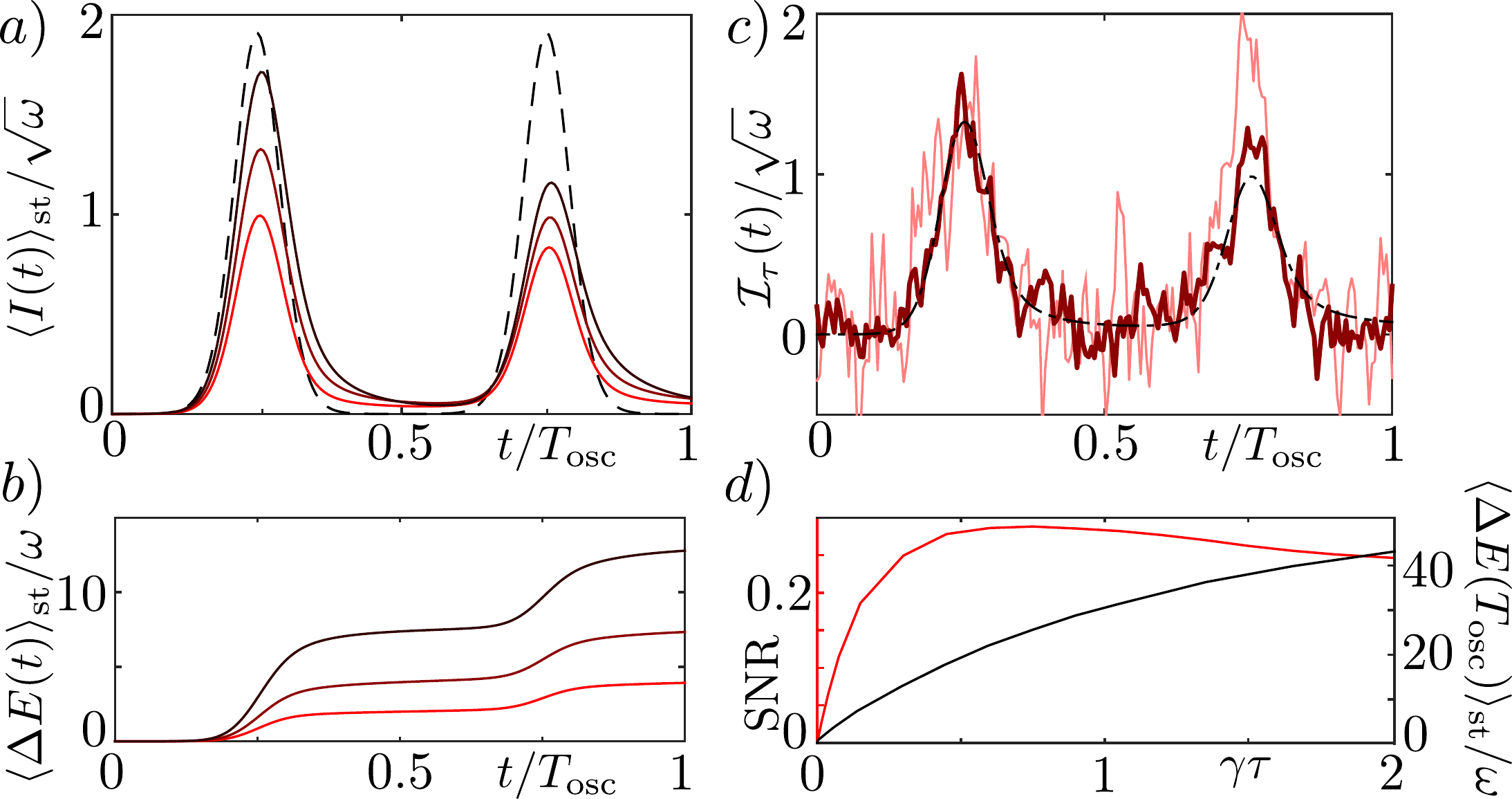}
\caption{Monitoring oscillations of a coherent wave packet in a HO ($\alpha=2$)
with a microscope at $z_{0}=0$ and $\sigma=0.3\ell_{0}$. (a) The
ensemble-averaged $\langle I(t)\rangle_{{\rm st}}$ over the oscillation
period $T_{{\rm osc}}=2\pi/\omega$ with increasing $\gamma/\omega=1$,
$2$, and $4$ (light to dark). Dashed line indicates the ideal transit
signal with no measurement (a.~u.). (b) Heating of the atom during
measurements. (c) Filtered homodyne current for $\gamma=2\omega$,
averaged over $50$ (thin) and $300$ (thick) measurements. (d) SNR at the
first peak ($t=T_{{\rm osc}}/4$) for a single measurement and the heating
for different $\gamma\tau$, with $\tau$ the filter integration time
(see text).}
\label{fig:fig3} 
\end{figure}

We consider below two modes of operation. In the first, the microscope
is placed at a given $z_{0}$, and we wish to `record a movie' of
the time dynamics of an atomic wave packet (e.g. a coherent state)
passing (repeatedly) through the observation zone. This requires a
time resolution better than the oscillation period, and corresponds
to the \emph{bad cavity limit} $\kappa\gg\omega$, where according
to Eq.~(\ref{eq:SME_elimination}) the homodyne current as a function
of time mirrors directly the wave packet motion at $z_{0}$ (c.f.~Fig.~\ref{fig:fig3},
and discussion below). As the second case we consider the \emph{good
cavity limit} $\kappa\ll\omega$. Here the observed homodyne signal
traces the atomic dynamics at $z_{0}$ cavity-averaged over many oscillation
periods \footnote{This situation is reminiscent of the sideband resolved laser cooling
of trapped particles in the Lamb-Dicke limit where $\gamma\ll\omega$
with $\gamma$ optical pumping rate \cite{gardiner2015quantum}}. However, as we show below, in this regime \emph{a slow scan} of
the focal point $z_{0}\equiv z_{0}(t)$ across the spatial region
of interest will turn the microscope into an effective QND device,
which maps out the spatial density associated with a particular energy
eigenfunction of the trapped particle with resolution $\sigma$. This
will be discussed below in the context of Fig.~\ref{fig:figure1}c-e,
where a particle is prepared initially in a state $\tilde{\rho}(0)=\sum_{n}p_{n}\ket{n}\bra{n}$
(e.g.~a thermal state), and in the spirit of QND measurements
a \emph{single scan} with the microscope
first collapses the atomic state into a particular motional eigenstate,
and subsequently `takes a picture' of its
spatial density. This ability of a \emph{single
scan} to reveal the density of energy eigenfunctions is in contrast to the first case above, where
the measurement is inherently destructive and a good SNR is only obtained
with repeated runs of the experiment.

\emph{Bad cavity limit and time-resolved dynamics} \textendash{} In
Fig.~\ref{fig:fig3}a we plot the ensemble averaged homodyne current $\langle I(t)\rangle_{{\rm st}}$
for a microscope positioned at $z_{0}=0$, which monitors the periodic
motion of an atomic wave packet in the HO. The atom is initially prepared
in a coherent state $\ket{\alpha}$ displaced from trap center with
$|\alpha|\gg1$, and the microscope detects the transit of the wave
packet with velocity $v=\sqrt{2}\ell{}_{0}|\alpha|\omega$ through
the trap center at times $t=1/4$, $3/4T_{\text{osc}}$ etc., with
$T_{\text{osc}}=2\pi/\omega$ the oscillator period. The time dependence
of the homodyne current reveals the shape of the wave packet for the
given resolution $\sigma=0.3\ell_{0}$. Fig.~\ref{fig:fig3}a plots $\langle I(t)\rangle_{{\rm st}}=2\sqrt{\gamma}\,\text{Tr}\,\left\{ f_{z_{0}}(\hat{z})\tilde{\rho}(t)\right\} $
for increasing measurement strengths $\gamma$, with $\tilde{\rho}(t)\equiv\langle\tilde{\rho}_{c}(t)\rangle_{{\rm st}}$
obeying Eq.~\eqref{eq:SME_elimination}. For the given parameters,
Fig.~\ref{fig:fig3}a displays the ability of the homodyne current to faithfully
represent the temporal shape of the wave packet, and reveals the measurement
backaction with increasing $\gamma$ as a successive distortion of
the signal with time. Fig.~\ref{fig:fig3}b quantifies this backaction as an increase
of the mean energy of the oscillator with time.

The SNR associated with these measurements is shown in Figs.~\ref{fig:fig3}c-d.
We define the SNR as $\langle{\cal I}_{\tau}(t)\rangle_{{\rm st}}^{2}/\langle\delta{\cal I}_{\tau}^{2}(t)\rangle_{{\rm st}}$
with ${\cal I}_{\tau}(t)\equiv\int_{\tau} I(t+t')dt'/\sqrt{\tau}$ the homodyne
current (\ref{eq:homodyneCurrentDef}) after a lowpass filter with
bandwidth $\tau^{-1}$, and the variance $\langle\delta{\cal I}_{\tau}^{2}(t)\rangle_{{\rm st}}\equiv\langle{\cal I}_{\tau}^{2}(t)\rangle_{{\rm st}}-\langle{\cal I}_{\tau}(t)\rangle_{{\rm st}}^{2}$.
We choose an integration
time $\tau$ sufficiently long to suppress the shot noise, but short
enough to resolve the temporal shape of the wave packet.
An optimal $\tau$ is related to the microscope spatial resolution, $\tau\sim\sigma/v=(\sigma/\ell_{0})\tau_{{\rm tr}}$
with $\tau_{\text{tr}}$ the transit time of the wave packet through
the focal region. In Fig.~\ref{fig:fig3}c we show
the homodyne current ${\cal I}_{\tau}(t)$ averaged over an increasing
number of measurements, and the convergence to the results of Fig.~\ref{fig:fig3}a.
In Fig.~\ref{fig:fig3}d the SNR in a single scan is plotted vs.~the
measurement strength $\gamma$. It shows the general behavior of non-QND
measurements~\cite{Clerk2010}: For small $\gamma$,
the SNR grows with increasing $\gamma$ due to suppression of the
shot noise. For large $\gamma$, SNR eventually drops down as the
measurement backaction induces strong additional noises.

\begin{figure}[t]
\centering{}\includegraphics[width=0.46\textwidth]{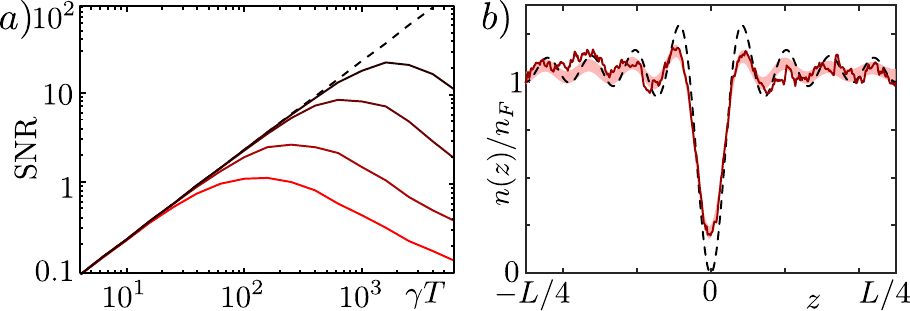}
\caption{Single-run scans in the QND regime. (a) SNR vs. $\gamma T$ for a
scan of an atom initialized in the state $\ket{1}$ of an HO for $\kappa/\omega=10$,
$1$, $0.25$, $0.1$ (light to dark), compared to an ideal QND measurement
(dashed line) for $\sigma=0.3\ell_{0}$ and $L=8\ell_{0}$. SNR is
taken at $z_{0}(t)=-\ell_{0}$ (theoretical maximum). (b) Scan of Friedel
oscillations for $N=16$ non-interacting fermions in a box of length
$L$ due to an impurity at $z=0$: The scanning signal (solid
line), the total noise variance (shaded area), and the theoretical
density profile $n(z)=n_{F}[1-\sin(2k_{F}z)/(2k_{F}z)]$ (dashed line)
with $n_{F}=k_{F}/\pi=N/L$.}
\label{fig:fig4} 
\end{figure}

\emph{Good cavity limit as emergent QND measurement }\textendash{}
A QND measurement requires that the associated observable commutes
with the system Hamiltonian. While $f_{z_{0}}(\hat{z})$ does not
commute with $\hat{H}_{\text{sys}}$, an effective QND measurement
emerges in the good cavity limit $\kappa\!\ll\!\omega$. We can see
this by transforming the SME (\ref{eq:SME_general}) to an interaction
picture with respect to $\hat{H}_{\text{sys}}$. This transformation results
in the replacement $f_{z_{0}}(\hat{z})\rightarrow\sum_{\ell}\hat f_{z_{0}}^{(\ell)}e^{-i\ell\omega t}$,
where $\hat f_{z_{0}}^{(\ell)}=\sum_{n}f_{n,n+\ell}|n\rangle\langle n+\ell|$
with $f_{mn}=\langle m|f_{z_{0}}(\hat{z})|n\rangle$. In a homodyne
measurement, where the current $I(t)$ is monitored with time resolution
$1/\kappa$, as filtered by the cavity, the terms rapidly oscillating
with frequencies $\ell\omega$ (motional sidebands) will not be resolved.
Thus homodyne detection provides a continuous measurement of $\hat f_{z_{0}}^{(0)}=\sum_{n}f_{n,n}|n\rangle\langle n|$ representing the emergent QND observable~\cite{newcomment}.


A formal derivation of these results is provided in Appendix~\ref{app:adiabatic_elimination} starting from
the SME (\ref{eq:SME_general}). There we derive for the homodyne
current $dX_{\phi}(t)\equiv I(t)dt=2\sqrt{\gamma}\,\langle \hat f_{z_{0}}^{(0)}\rangle_{c}+dW(t)$
with $\langle\dots\rangle_{c}={\rm Tr}\{\ldots\tilde{\rho}_{c}(t)\}$, where
the conditional density operator $\tilde{\rho}_{c}(t)$ obeys the SME 
\begin{align}
d\tilde{\rho}_{c} & =-\frac{i}{\hbar}[\hat{H}_{{\rm sys}},\tilde{\rho}_{c}]\,dt+\sum_{\ell\neq0}\frac{\gamma}{1+(2\omega\ell/\kappa)^{2}}{\cal D}[\hat f_{z_{0}}^{(\ell)}]\tilde{\rho}_{c}\,dt\nonumber \\
 & \quad+\gamma{\cal D}[\hat f_{z_{0}}^{(0)}]\tilde{\rho}_{c}\,dt+\sqrt{\gamma}{\cal H}[\hat f_{z_{0}}^{(0)}]\tilde{\rho}_{c}\,dW(t),\label{eq:SME_good_cavity_elimination-1}
\end{align}
with $\gamma$ the measurement strength defined above 
(assuming resonant driving). To provide a physical interpretation,
we take matrix elements of Eq.~(\ref{eq:SME_good_cavity_elimination-1})
in the energy eigenbases and obtain a (nonlinear) \emph{stochastic
rate equation }(SRE) for the trap-state populations $p_{n}=\langle n|\tilde{\rho}_{c}|n\rangle$:
\begin{align}
dp_{n} & =\frac{\gamma}{1+(2\omega/\kappa)^{2}}\left[A_{n}^{(+)}p_{n+1}+A_{n}^{(-)}p_{n-1}-B_{n}p_{n}\right]dt\nonumber \\
 & \quad+2\sqrt{\gamma}p_{n}\left[f_{nn}-\sum\nolimits _{m}f_{mm}p_{m}\right]dW(t).\label{eq:diagonal_SME-1}
\end{align}
Here $A_{n}^{(\pm)}\equiv|f_{n,n\pm1}|^{2}$, $B_{n}\equiv A_{n}^{(+)}+A_{n}^{(-)}$,
and for simplicity we have kept only the dominant terms $\ell=0,\pm1$ for $\kappa/\omega\ll1$. We emphasize that Eq.~(\ref{eq:diagonal_SME-1})
involves two time scales. The stochastic term in the second line describes
the collapse of the density operator to a particular trap eigenstate
$\tilde{\rho}_{c}(t)\rightarrow\ket{n}\bra{n}$ within a collapse
time $T_{\text{coll}}\sim1/\gamma$. In contrast, the first line is
a redistribution of population between the trap levels, for a much longer
dwell time $T_{\text{dwell}}\sim(2\omega/\kappa)^{2}\gamma^{-1}\gg T_{\text{coll}}$.
As a result, the time evolution consists of a rapid collapse to an
energy eigenstate $\ket{n}$, followed by a sequence of rare quantum
jumps $n\rightarrow n\pm1$ on the time scale $T_{\text{dwell}}$.
The QND mode of the microscope exploits these two time scales by scanning
the focal point across the system, $-L/2<z_{0}(t)<L/2$, in a time $T_{\text{coll}}\ll T\lesssim T_{\text{dwell}}$. Starting the measurement scan, the motional state will first collapse to a particular
state $\ket{n}$, with the subsequent scan revealing the spatial density
profile $\langle n|\hat f_{z_{0}}^{(0)}|n\rangle=\int dz\,f_{z_{0}}(z)|\langle z|n\rangle|^{2}$.

Fig.~\ref{fig:figure1}c shows a simulation representing a \textit{single
run} in the QND regime ($\kappa/\omega=0.1$) based on integrating
the SME~(\ref{eq:SME_good_cavity_elimination-1}). The atom at $t=0$
is prepared in a thermal motional state of the HO, $\tilde{\rho}(0)=\sum_{n}p_{n}\ket{n}\bra{n}$
with $n_{\rm th}=0.6$. We perform three consecutive spatial scans covering
$-L/2<z_{0}(t)<L/2$ ($L=10\ell_{0}$), each in a time interval $T$
($\gamma T=5000$). For the run shown in Figs.~\ref{fig:figure1}c-e,
the QND measurement in \textit{scan 1} first projects the atomic trap
population into $\ket{0}$, followed by a transition to $\ket{1}$
at time $t_{1}$, and $\ket{1}\rightarrow\ket{0}$ at $t_{2}$ in
\textit{scan 2}, and no transition in \textit{scan 3}. The homodyne
current ${\cal I}_{\tau}(t)$ associated with these single scans is
a faithful representation of the spatial density distributions of
eigenfunctions $|\langle z|n\rangle|^{2}$. In Fig.~\ref{fig:fig4}a the SNR of single scans of a pure state is shown against  the (dimensionless) measurement strength $\gamma T$. By decreasing $\kappa/\omega$ we greatly suppress the measurement back-action, rendering them into rarer quantum jumps, thus improving the SNR.

The concept of a scanning microscope to observe \textit{in vivo} cold
atom dynamics is readily adapted to a quantum many-body system, and we show in Fig.~\ref{fig:fig4}b a single spatial
scan of the Friedel oscillation of an non-interacting Fermi sea in
the presence of a single impurity~(see Appendix~\ref{app:friedel} for details). 
While we have focused on homodyne
measurement in CQED for continuous readout (with experimental feasibility discussed in Appendix~\ref{app:feasibility}), atomic physics setups
provide interesting alternative routes to achieve weak continuous
measurement, e.g.~coupling to atomic ensembles via Rydberg interactions~\cite{SaffmanRMP,Labuhn2016,Bernien:2017aa}.

\begin{acknowledgments}
We acknowledge discussions with P.~Hauke, and thank P. Grangier, I.B. Mekhov, K. M\o{}lmer and L. Orozco for useful comments on the manuscript. Work at Innsbruck is supported
by the Austrian Science Fund SFB FoQuS (FWF Project No. F4016-N23) and 
the European Research Council (ERC) Synergy Grant UQUAM. 
\end{acknowledgments}

\bibliography{scanningMicroscope_bib}

\begin{thebibliography}{52}
\bibitem [{com()}]{comment1}%
  \BibitemOpen
  \href@noop {} {}\bibinfo {note} {Note that the local density operator does
  not commute with the Hamiltonian. Nevertheless, in our setup, we can measure
  local densities for energy eigenstates with no back-action --- an emergent
  QND measurement, see text and \cite{newcomment}.}\BibitemShut {Stop}%
\bibitem [{new()}]{newcomment}%
  \BibitemOpen
  \href@noop {} {}\bibinfo {note} {The concept of emergent QND observable and
  measurement can be defined in a general context, as detailed in
  Appendix~\ref{app:eQND}.}\BibitemShut {Stop}%
\end{thebibliography}

\newpage
\appendix
\section{Quantum non-Demolition vs. \textit{Emergent} Quantum non-Demolition Measurements}
\label{app:eQND}

In this section we summarize the concept of emergent quantum non-demolition (QND) measurements in a more formal way. 

A familiar definition of a QND measurement \cite{Braginsky1992,Clerk2010}  considers an observable ${\hat A}$ to be QND, if it  commutes with the system Hamiltonian $[\hat H, {\hat A}]=0$. Such a QND observable can be continuously measured with an arbitrary high Signal-to-Noise Ratio (SNR)~\cite{Braginsky1992,Clerk2010}. 

In general, for a system observable $\mathcal{\hat O}$, which does not commute with the Hamiltonian, $[\hat H, \mathcal{\hat O}]\neq0$, we define as emergent QND observable 
\begin{align}
\hat{\mathcal{O}}_{\rm eQND} \equiv \sum_n \ket{n}\bra{n}\hat{\mathcal{O}}\ket{n}\bra{n},
\end{align}
with $\ket{n}$ the energy eigenstates. Measurement of $\hat{\mathcal{O}}_{\rm eQND}$ provides the same information as of $\hat{\mathcal{O}}$ for energy eigenstates, but in a non-destructive way. This enables studying properties of energy eigenstates of various quantum systems with very high precision and from different perspectives provided by the corresponding observables $\mathcal{\hat O}$. The emergent QND measurement in context of the quantum scanning microscope, as discussed in the main text, considers the eQND observable defined from the $\delta$-like probe $f(\hat z)$, where $\hat z$ is the position operator. This allows in particular to map out atomic densities of energy eigenstates for harmonic oscillator and Freidel oscillations for many-body systems in a single scan with high SNR, as illustrated in Figs.~1 and 4 of the main text.


\section{Engineering of The Sub-wavelength Focusing Function $\phi_{z_0}(z)$}
\label{app:FocusFunction}
Here we discuss in detail the realization of the focusing function $\phi_{z_0}(z)$  [c.f. Eq.~(1) of the main text], showing that subwavelength resolution can be achieved along with negligible additional forces on the atom.

\subsection{Sub-wavelength spin structure with negligible non-adiabatic potential}
The atomic internal levels for implementing the focusing function, shown in Fig.~2a of the main text, consists of a $\Lambda$-system formed by $|g\rangle,|r\rangle,|e\rangle$, described by the Hamiltonian
\be
\label{eq:4levelSystem}
\begin{split}
\hat{H}_{ \mathrm{a} } \!= &\!-\!\!\hbar \left(\Delta_e\!+\! i\frac{\Gamma_e}{2}\right)\!\hat{\sigma}_{ee} 
 + \frac{\hbar }{2} \left[ \Omega_0(z)\hat{\sigma}_{eg} \!+\! \Omega_1(z)\hat{\sigma}_{er} \!+\! \mathrm{H.c.} \right],
\end{split}
\ee
where $\Gamma_e$ is the decay rate of the excited state, and we assume Raman resonance $\Delta_r=0$.
Diagonalizing $\hat{H}_{ \mathrm{a} }$ gives the eigenstates
\be
\begin{split}
\ket{D(z)} &= \sin\theta (z)\ket{g}-\cos\theta (z) \ket{r},\\
\ket{+(z)} &=\cos\chi (z) \ket{e} + \sin\chi (z) [\cos\theta (z) \ket{g}+\sin\theta (z)\ket{r}],\\
\ket{-(z)} &=\sin\chi (z) \ket{e} - \cos\chi (z)[\cos\theta (z)\ket{g}+\sin\theta (z)\ket{r}],
\end{split}
\ee
with the corresponding eigenenergies $E_{D}=0$ and $E_{\pm} (z)=-({\hbar }/2)\{\tilde{\Delta}_{e}\mp[\Omega_0^2(z)+\Omega_1^2(z)+\tilde{\Delta}_e^2]^{1/2}\}$, where $\tilde{\Delta}_e=\Delta_e+i\Gamma_e/2$,
and the mixing angles defined by $\theta(z)=\arctan[\Omega_1(z)/\Omega_0(z)]$ and $\chi(z)=-({1}/{2})\arctan[\sqrt{\Omega_0^2(z)+\Omega_1^2(z)}/\tilde{\Delta}_e]$.  We note that the \emph{dark state} $\ket{D(z)}$ is decoupled from the dissipative excited state $\ket{e}$, and its spin structure is varying in space controlled by the Rabi frequency configuration.

We are interested in the regime where $\textrm{Re}[E_{\pm}(z)]$ is much larger than the other energy scales in the model. In the spirit of the Born-Oppenheimer (BO) approximation, we study the \emph{slow dynamics} by assuming the atomic internal state remains in $\ket{D(z)}$ adiabatically. This allows us to design the desired sub-wavelength spin structure $|\langle r| D(z)\rangle|^2=\cos^2\theta$. Such a spatially varying internal spin is, however, necessarily accompanied by non-adiabatic corrections to the atomic external motion~\cite{Lacki2006,Gorshkov2016}
\be
\label{Vna}
V_{ na} (z)= \bra{D(z)}\frac{\hat{p}_z^2}{2m}\ket{D(z)}=\frac{\hbar^2}{2m}[\partial_z\theta(z)]^2.
\ee

We now show that $|\langle r|D(z)\rangle|^2$ can be made nano-scale with negligible $V_{na}(z)$. We consider the Rabi frequencies
\be
\label{Omega1Expression}
\begin{split}
\Omega_{0}(z)&=\epsilon \Omega_c,\\
\Omega_1(z)& = \Omega_c [ 1 + \beta - \cos k_1 (z-z_0) ],
\end{split}
\ee
where $\Omega_c$ is a large reference frequency (assumed real positive) and $0<\epsilon\sim\beta\ll1$. Physically, $\Omega_1(z)$ can be realized, e.g., by super-imposing three phase-coherent laser beams where the first two lasers form the standing wave $\Omega_c\cos k_1(z-z_0)$, and the third propagates perpendicularly, to provide the offset $\Omega_c(1+\beta)$.

\begin{figure}[t]
\centering{}\includegraphics[width=0.25\textwidth]{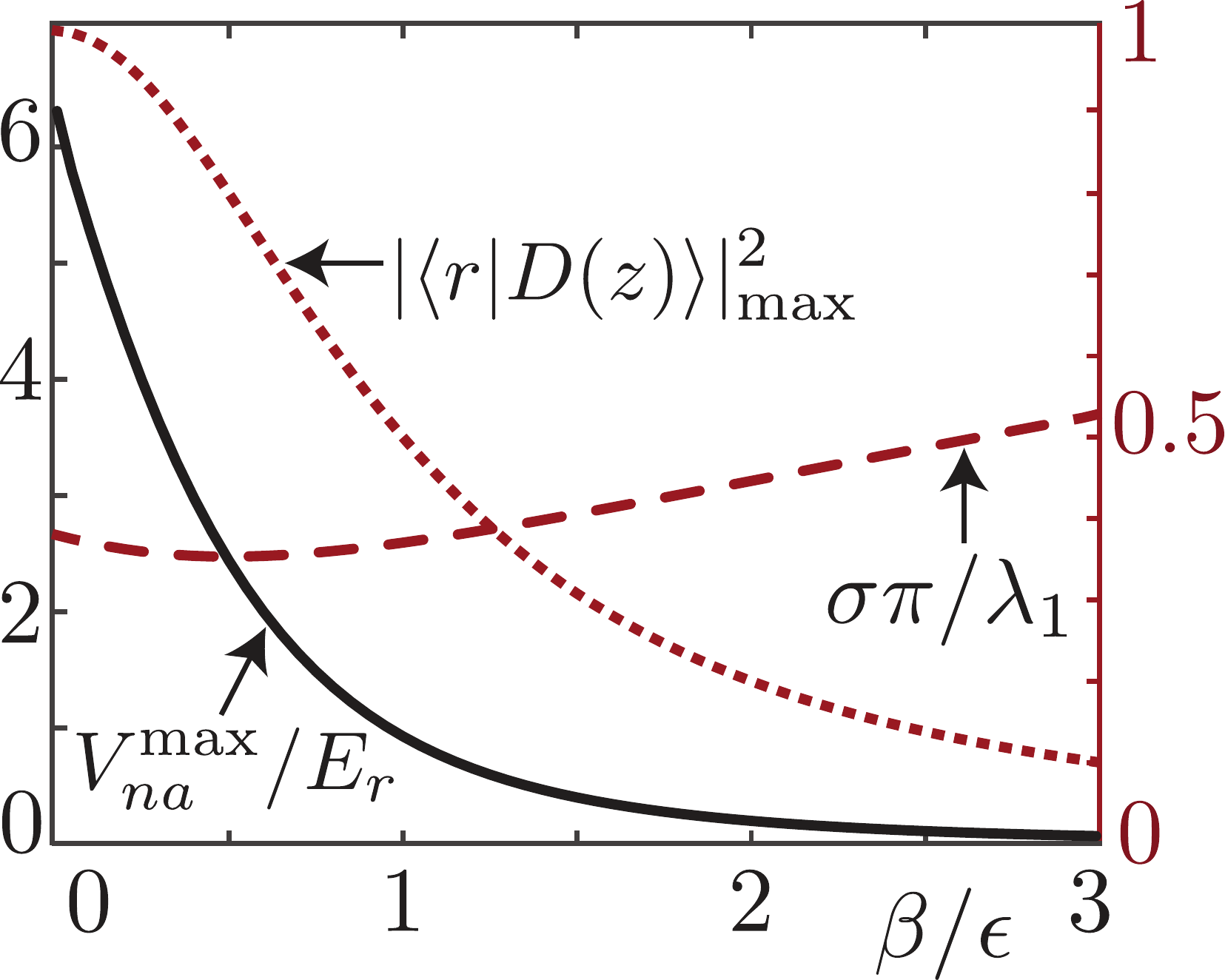}
\caption{The resolution $\sigma$ and the maximum of the non-adiabatic potential
$V_{na}(z)$ {[}c.f., Eq.~\eqref{Vna}, in unit of the recoil energy
$E_{r}=\hbar^{2}k_{1}^{2}/2m${]}, vs.~$\beta/\epsilon$, for the
laser configuration Eq.~\eqref{Omega1Expression}. Also shown is
the maximum overlap between $\ket{D(z)}$ and $\ket{r}$. Parameters:
$\epsilon=0.1$. We note that $V_{na}(z)$ is strongly suppressed
for $\beta/\epsilon\gg1$.}
\label{fig:fig1_supp} 
\end{figure}

For Rabi frequencies in Eq.~\eqref{Omega1Expression}, the resolution $\sigma$, quantified by the full width at half maximum (FWHM) of $|\langle r|D(z)\rangle|^2$, is given in the limit $\epsilon\ll 1$ by 
\be
\label{appResolution}
\sigma=\frac{\sqrt{2}\lambda_1}{\pi}\Big(\sqrt{\epsilon^2+2\beta^2}-\beta\Big)^{1/2},
\ee
with $\lambda_1=2\pi/k_1$. The non-adiabatic potential is
\be
\label{nonAdiabaticCorrection}
V_{na}(z)=\frac{\hbar^2k^2_1}{2m \epsilon^2}\left(\frac{\sin k_1(z-z_0)}{1+[1+\beta-\cos k_1(z-z_0)]^2/\epsilon^2}\right)^2.
\ee
Importantly, $V_{na}(z)$ decreases rapidly by increasing the ratio $\beta/\epsilon$, as shown in Fig.~\ref{fig:fig1_supp}. Physically, increasing $\beta/\epsilon$ reduces the maximal population transfer onto the state $\ket{r}$ during the adiabatic motion, $|\langle r|D(z)\rangle|^2_{\rm max} = (1+\beta^2/\epsilon^2)^{-1}$, thus suppressing the corresponding non-adiabatic potential. This sub-wavelength spin structure, with negligible $V_{na}(z)$, is exploited to realize the focusing function $\phi_{z_0}(z)$, as we show below.

\subsection{Sub-wavelength focusing function $\phi_{z_0}(z)$}
Being a part of the $\Lambda$-system, the level $\ket{r}$ is also
coupled to another level $\ket{t}$ through a cavity mode $\hat{c}$,
resulting in an effective dispersive coupling $\hat{H}_{{\rm ac}}=\hbar g(\hat{z})^{2}\hat{c}^{\dag}\hat{c}/\Delta_{t}$
between $\ket{r}$ and the cavity mode (the effects of the spontaneous
decay of the state $\ket{t}$ will be discussed below). After projecting
onto the dark state $\ket{D(z)}$ in the BO approach, one obtains
the desired sub-wavelength atom-cavity coupling 
\begin{equation}
\hat{H}_{{\rm coup}}=\frac{\hbar g^{2}(z)}{\Delta_{t}}|\langle r|D(z)\rangle|^{2}\hat{c}^{\dag}\hat{c}\equiv\phi_{z_{0}}(z)\hat{c}^{\dag}\hat{c},\label{derivationHcoup}
\end{equation}
with the spatial resolution given by Eq.~\eqref{appResolution} {[}notice
that $g(z)$ varies slowly on the scale $\sigma${]}.

As mentioned in the main text, it is convenient to write the focusing
function in the form $\phi_{z_{0}}(z)\equiv{\cal A}f_{z_{0}}(z)$,
where $\mathcal{A}$ has the dimension of energy and $f_{z_{0}}(z)$
is dimensionless and normalized. We choose the normalization $\int dzf_{z_{0}}(z)=\ell_{0}$
with $\ell_{0}$ being the characteristic length scale of the system
under measurement, such that the matrix elements of $f_{z_{0}}(\hat{z})$
are of order~$1$.

Note that through this coupling, the stationary coherent field inside
the driven cavity exerts a force on the atom, $V_{{\rm OL}}(z)=\hbar g^{2}(z)|\alpha|^{2}|\langle r|D(z)\rangle|^{2}/\Delta_{t}$,
where $\alpha=\sqrt{\kappa}\mathcal{E}(i\delta-\kappa/2)^{-1}$ is
the amplitude of the stationary field, $\delta$ and $\mathcal{E}$
are the detuning and the strength of the cavity driving laser, respectively.
This force can be compensated by simply detuning the Raman resonance
with the offset $\Delta_{r}=g^{2}(z_{0})|\alpha|^{2}/\Delta_{t}$
(c.f. Fig.~2a in the main text), which results in nearly perfect
compensation of $V_{{\rm OL}}(z)$ for the dark state $\ket{D(z)}$.

We also note that the focusing function $\phi_{z_{0}}(z)$ in by Eq.~\eqref{Omega1Expression}
has a periodic set of peaks separated by $\lambda_{1}=2\pi/k_{1}$.
To design a single-peak $\phi_{z_{0}}(z)$ one can simply choose a
spatially dependent $\Omega_{0}(z)$ which is tightly ($\sim\lambda_{1}$)
focused at position $z_{0}$.

\subsection{Spontaneous emission}
\label{app:sponE}
Here we discuss the spontaneous emission of the
dark state originated from the spontaneous decay of the states $\ket{e}$
and $\ket{t}$ entering the construction of the focusing function
(see Fig.~2a of the main text). A more formal derivation of the same results using the
stochastic master equation including the atomic internal states and
the associated spontaneous decay terms will be presented in a follow-up
paper~\cite{pra}.

The spontaneous decay rate of the state $\ket{e}$ enters through
the residual coupling between the dark state $\ket{D(z)}$ and the
bright states $\ket{\pm(z)}$ in the $\Lambda$-configuration, due
to the atomic kinetic term. As shown in \cite{Lacki2006}, the corresponding decay rate of $\ket{D(z)}$
scales with the Rabi frequencies as $\propto[\Omega_{1}^{2}(z)+\Omega_{0}^{2}(z)]^{-1}$,
and can be strongly suppressed by choosing large Rabi frequencies.

The spontaneous decay of $\ket{D(z)}$ due to virtual population of
the state $\ket{t}$ (resulting from $\hat{H}_{{\rm coup}}$) is the
dominant decay channel, and the corresponding decay
rate can be calculated as
\begin{equation*}
\gamma_{D}(z)=\frac{g^{2}(z)}{\Delta_{t}^{2}}\left(\frac{4\mathcal{E}^{2}}{\kappa}\right)\Gamma_{t}|\langle r|D(z)\rangle|^{2}=\gamma_{{\rm sp}}f_{z_{0}}(z),
\end{equation*}
where $4\mathcal{E}^{2}/\kappa$ is the mean photon number in the
driven cavity, $\Gamma_{t}$ is the spontaneous emission rate of the
state $\ket{t}$, and we introduce the average spontaneous decay rate
\begin{equation*}
\gamma_{{\rm sp}}=\frac{1}{\ell_{0}}\int dz\gamma_{D}(z)=4\mathcal{A}\frac{\mathcal{E}^{2}\Gamma_{t}}{\hbar\kappa\Delta_{t}}.
\end{equation*}
This has to be compared with the measurement strength $\gamma=[4\mathcal{{\cal A}E}/(\hbar\kappa)]^{2}$, with the result
\be
\begin{split}
\frac{\gamma}{\gamma_{{\rm sp}}}=\frac{4\mathcal{{\cal A}}\Delta_{t}}{\hbar\kappa\Gamma_{t}}&=\frac{4}{\kappa\Gamma_{t}\ell_{0}}\int dzg^{2}(z)|\langle r|D(z)\rangle|^{2}\\&\simeq 4\mathcal{C}\frac{\sigma}{\ell_{0}}|\langle r|D(z)\rangle|_{{\rm max}}^{2},
\end{split}
\label{eq: measurementVSsp}
\ee
where $\mathcal{C}\equiv g^{2}(z_{0})/(\kappa\Gamma_{t})$ is the
cavity cooperativity, and we use the approximation $\int dzg^{2}(z)|\langle r|D(z)\rangle|^{2}\simeq\sigma g^{2}(z_{0})|\langle r|D(z)\rangle|_{{\rm max}}^{2}$.

An implementation of the proposed microscope would require $\gamma\gg\gamma_{{\rm sp}}$,
so that large SNR can be achieved during the time when spontaneous
emission is still negligible. This condition can be met with today's
high-Q optical cavities, as shown below. 

\section{Experimental feasibility}
\label{app:feasibility}
In this section we show that the proposed microscope can be implemented
in the state-of-the-art experiments involving cold atoms/trapped
ions and optical cavities, and discuss typical experimental parameters.

First, as discussed in Appendix~\ref{app:sponE}, a prerequisite
for the operation of the microscope is $\gamma\gg\gamma_{{\rm sp}}$.
The cooperativity $\mathcal{C}$ of high-Q optical cavity can exceed
100 in state-of-the-art experiments ~\cite{Hood98}. To make an estimation
we choose $\mathcal{C}=150$, $\sigma=0.3\ell_{0}$ and $|\langle r|D(z)\rangle|_{{\rm max}}^{2}=0.4$
(which suffices to render $V_{na}$ being negligible, c.f. Fig.~\ref{fig:fig1_supp}),
yielding $\gamma/\gamma_{{\rm sp}}\simeq75$. Such a high ratio guarantees
that spontaneous emission is indeed negligible for the observation
of the key predictions in the main text: for $\gamma T\simeq75$ with
$T$ being the total measurement time, one gets ${\rm SNR}\gg1$ (c.f.
Fig.~4a of the main text) for a single scan of atomic motional eigenstates
in the QND mode of the microscope.

Second, the two operation modes of the microscope require either $\omega\ll\kappa$
or $\omega\gg\kappa$. While the first region $\omega\ll\kappa$ us naturally obtained using cavities with a sufficiently large linewidth, the second condition is also realistic. For example, Ref.~\cite{Wolke75}
reports a coupled BEC-cavity setup with $\kappa\simeq2\pi\times4.5{\rm kHz}$
which is far smaller than the recoil energy of light-mass alkalies
(e.g., $E_{r}\simeq 2\pi\times60{\rm {kHz}}$ for $^{7}$Li at the D$2$
line). Trapped ions provides another platform for reaching the good
cavity limit, due to their large oscillation frequency ($\sim$MHz).

\section{Perturbative Elimination of the Cavity Field}
\label{app:adiabatic_elimination}
In this section we consider the relation between the homodyne current $I(t)$ and the localised microscope probe $f_{z_0}(\hat z)$. To obtain the connection we eliminate the cavity field from the stochastic dynamics described by the Eq.~(3) in the main text. The aim is to derive effective stochastic master equations (5) and (6) in the main text, with corresponding photocurrents in `bad' and `good' cavity limits respectively.

To make the discussion more general, first, we consider an arbitrary system which is coupled to the cavity field $\hat c$ via interaction $\hat H_{{\rm int}}=\hbar \varepsilon \hat f(\hat c+\hat c^{\dagger})$ where $\hat f$
is a system operator and the coupling is assumed to be weak compared to the cavity
decay rate $\varepsilon\ll\kappa$. This model is related to the atomic system from the main text coupled to a driven cavity via the interaction Eq.~(1) linearised around the steady state intracavity field.

Transforming to an interaction picture with respect to the system
Hamiltonian $\hat H_{{\rm sys}}$ we obtain the following SME describing
the dynamics of the full setup under continuous homodyne monitoring
of the cavity output field:
\begin{align}
d\rho_c&=-i[\varepsilon \hat f(t)(\hat c+\hat c^{\dagger})+\delta \hat c^{\dagger}\hat c,\rho_c]dt+\kappa\mathcal{D}[\hat c]\rho_c dt\notag\\
&\quad+\sqrt{\kappa}\mathcal{H}[\hat ce^{-i\phi}]\rho_c dW(t),\label{eq:full_SME}
\end{align}
where $\delta$ is the cavity detuning, $\phi$ is the homodyne angle,
and $\hat f(t)=e^{i\hat H_{{\rm sys}}t}\hat fe^{-i\hat H_{{\rm sys}}t}$. The corresponding
homodyne current reads:
\begin{equation}
dX_{\phi}(t) \equiv I(t)dt=\sqrt{\kappa}\left\langle \hat c e^{-i\phi} + \hat c^{\dagger}e^{i\phi}\right\rangle _{c}dt+dW(t)\label{eq:app_photocurrent}
\end{equation}
where $\langle\dots\rangle_{c}\equiv\mathrm{Tr}\{\ldots\,\rho_c(t)\}$
refers to an expectation value with respect to the conditional density
matrix.
We eliminate the cavity field along the lines of \cite{Cernotik2015}.
First, we simply trace out the cavity dynamics from the SME~\eqref{eq:full_SME}
to obtain a stochastic equation for the system density matrix only
($\tilde\rho_{c}={\rm Tr}_{T}\rho_c$):
\begin{align}
d\tilde\rho_{c} & =-i\varepsilon\left[\hat f(t),\hat \eta+\hat \eta^{\dagger}\right]dt+\sqrt{\kappa}\left(\hat \mu e^{-i\phi}+\hat \mu^{\dagger}e^{i\phi}\right)dW(t)\label{eq:rho_S}
\end{align}
where we define operators $\hat \eta={\rm Tr}_{T}\{\hat c\rho_c\}$ and $\hat \mu=\hat \eta-\langle \hat c\rangle \tilde\rho_{c}$
such that $\langle \hat c\rangle ={\rm Tr}_{S}\,\hat \eta$, and operations ${\rm Tr}_{T}$ and ${\rm Tr}_{S}$ stand for the partial traces over states of the cavity ($T$ for transducer) and the system respectively. We derive an effective equation for $\tilde\rho_{c}$ up to the second order in the perturbation
$\varepsilon$ for the deterministic term and up to a linear stochastic
term: $d\tilde\rho_{c}=O(\varepsilon^{2})dt+O(\varepsilon)dW(t)$. This restricts
equations for the operators $\hat \eta$ and $\hat \mu$ to $d\hat \eta(d\hat \mu)=O(\varepsilon)dt+O(1)dW(t)$.
The equation of motion for $\eta$ operator reads
\begin{align}
d\hat \eta & ={\rm Tr}_{T}\{\hat c\,d\rho_c\}\nonumber \\
 & =-i\varepsilon{\rm Tr}_{T}\left\{ \hat c\left[\hat f(t)(\hat c+\hat c^{\dagger}),\rho_c\right]\right\} dt-\left(\frac{\kappa}{2}-i\delta\right)\hat \eta \,dt\nonumber \\
 & \quad+\sqrt{\kappa}\,{\rm Tr}_{T}\left\{ \hat c(\hat c-\langle \hat c\rangle) {\rho}_c e^{-i\phi}+\hat c(\hat c^{\dagger}-\langle \hat c^{\dagger}\rangle) {\rho}_ce^{i\phi}\right\} dW(t)\nonumber \\
 & \simeq-i\varepsilon \hat f(t)\tilde\rho_{c}dt-\left(\frac{\kappa}{2}-i\delta\right)\hat \eta dt,\label{eq:eta}
\end{align}
where in the first deterministic term and in the stochastic term we used
the fact that $\rho_c=\tilde\rho_{c}\otimes\rho_{T}$ to zeroth order in $\varepsilon$
and that the unperturbed cavity is in a vacuum steady state such that $\langle \hat c\hat c^{\dagger}\rangle =1$,
$\langle \hat c\hat c\rangle =\langle \hat c^{\dagger}\hat c\rangle =0$.
Next, for the cavity mean field we have:
\begin{align}
d\langle \hat c\rangle  & ={\rm Tr}_{S}d\hat \eta\nonumber \\
 & \simeq-i\varepsilon\langle \hat f(t)\rangle dt-\left(\frac{\kappa}{2}-i\delta\right)\langle \hat c\rangle dt\label{eq:app_mean_c}
\end{align}
This equation is first order in $\varepsilon$ which means, to define operator $\hat \mu$, we need to know $\tilde\rho_{c}$ to the zeroth order in $\varepsilon$. It is constant in this approximation ($d\tilde\rho_{c}=0$) and we obtain an equation for the $\hat \mu$ operator using It\^{o} rule: 
\begin{align}
d\hat \mu & =d\hat \eta-\left\{ \left(d\langle \hat c\rangle \right)\tilde\rho_{c}+\langle \hat c\rangle d\tilde\rho_{c}+\left(d\langle \hat c\rangle \right)d\tilde\rho_{c}\right\} \nonumber \\
 & \simeq-i\varepsilon\left\{ \hat f(t)-\langle \hat f(t)\rangle \right\} \tilde\rho_{c}dt-\left(\frac{\kappa}{2}-i\delta\right)\hat \mu\, dt\label{eq:mu}
\end{align}
Plugging the solutions of the Eqs.~\eqref{eq:eta} and \eqref{eq:mu}
into the equation of motion for the system density operator~\eqref{eq:rho_S}
we recover an effective equation with the necessary precision in $\varepsilon$.
There are two cases to consider.

\emph{`Bad' cavity} \textemdash{} If the free evolution of the system
can be neglected on a time scale of the cavity decay $1/\kappa$ (for
our harmonic oscillator $\kappa\gg\omega$) we obtain:
\begin{equation*}
\hat \eta \simeq-i\varepsilon\frac{\hat f(t)}{\kappa/2-i\delta}\tilde\rho_{c},\qquad
\hat \mu \simeq-i\varepsilon\frac{\hat f(t)-\langle \hat f(t)\rangle }{\kappa/2-i\delta}\tilde\rho_{c}.
\end{equation*}
Substituting these expressions into the Eq.~\eqref{eq:rho_S} and restoring the Schr\"{o}dinger picture we arrive
at the effective SME:
\begin{equation}
d\tilde\rho_{c}=-\frac{i}{\hbar }[\hat H_{{\rm eff}},\tilde\rho_{c}]dt+\gamma\mathcal{D}[\hat f]\tilde\rho_{c}dt+\sqrt{\gamma}\mathcal{H}[\hat f]\tilde\rho_{c}dW(t)
\label{eq:app:bad_cavity_SME}
\end{equation}
where $\hat H_{{\rm eff}}=\hat H_{{\rm sys}}+(\hbar\, \delta\,\varepsilon^{2}\hat f^{2})/\{(\kappa/2)^{2}+\delta^{2}\}$
and $\gamma=(\varepsilon^{2}\kappa)/\{(\kappa/2)^{2}+\delta^{2}\}$.
The homodyne phase is chosen to maximize the signal in the photocurrent
($\phi=-{\pi}/{2}+{\rm arctan}\{{2\delta}/{\kappa}\}$) such that
\begin{equation}
dX_{\phi}(t)\equiv I(t)dt=2\sqrt{\gamma}\langle \hat f\rangle _{c}dt+dW(t)
\label{eq:app:bad_cavity_current}
\end{equation}
which
is obtained by substituting solution of Eq.~\eqref{eq:app_mean_c}
into Eq.~\eqref{eq:app_photocurrent}. In the main text we consider the cavity
driven by a coherent field $\mathcal{E}$ such that the coupling
coefficient is given by $\varepsilon=({\mathcal{AE}}/{\hbar })\{{\kappa}/(\kappa^2/4+\delta^{2})\}^{1/2}$. Defining $\hat f=f_{z_0}(\hat z)$ and setting zero detuning $\delta=0$ the equations \eqref{eq:app:bad_cavity_current} and \eqref{eq:app:bad_cavity_SME} become Eqs.~(4) and (5) in the main text. 

Ensemble averaging the stochastic dynamics \eqref{eq:app:bad_cavity_SME} over individual trajectories
results in the corresponding ME for the unconditional density matrix
$\tilde{\rho}(t)=\left\langle \tilde\rho_{c}(t)\right\rangle _{{\rm st}}$:
$$
\frac{d\tilde{\rho}}{dt}=-\frac{i}{\hbar }[\hat H_{{\rm eff}},\tilde{\rho}]+\gamma\mathcal{D}[\hat f]\tilde{\rho}.
$$
Here we used the non-anticipating property of the stochastic differential equation in It\^{o} form $\left\langle \dots dW(t)\right\rangle _{{\rm st}}=0$.

\emph{`Good' cavity} \textemdash{} In the case of harmonic oscillator
$\hat H_{{\rm sys}}=\hbar \omega(\hat a^{\dagger}\hat a+1/2)$ the system coupling operator (localised probe $f_{z_0}(\hat z)$)
in the interaction picture reads $\hat f(t)=\sum_{\ell}\hat f^{(\ell)}e^{-i\ell\omega t}$,
where $\hat f^{(\ell)}=\sum_{n}f_{n,n+l}|n\rangle\langle n+\ell|$ with
$f_{mn}=\langle m|f_{z_0}(\hat z)|n\rangle$. This allows one to integrate Eq.~\eqref{eq:eta}
and \eqref{eq:mu} assuming slow time dependence of $\tilde\rho_{c}$ as
follows:
\begin{align*}
\hat \eta & \simeq-i\varepsilon\sum_{\ell}\frac{\hat f^{(\ell)}e^{-i\ell\omega t}}{\kappa/2-i(\delta+\omega\ell)}\tilde\rho_{c}\\
\hat \mu & \simeq-i\varepsilon\sum_{\ell}\frac{\left\{ \hat f^{(\ell)}-\langle \hat f^{(\ell)}\rangle \right\} e^{-i\ell\omega t}}{\kappa/2-i(\delta+\omega\ell)}\tilde\rho_{c}
\end{align*}
Substituting the results into the Eq.~\eqref{eq:rho_S}, keeping only
non-rotating deterministic terms due to $\kappa \ll \omega$ in the `good' cavity limit (secular approximation), and transforming
back to the Schr\"{o}dinger picture we obtain:
\begin{align}
d\tilde\rho_{c} & \!=\!-\frac{i}{\hbar}[\hat H_{{\rm eff}},\tilde\rho_{c}]dt+\!\sum_{\ell}\frac{\varepsilon^{2}\kappa}{(\kappa/2)^{2}\!+\!(\delta\!+\!\omega\ell)^{2}}\mathcal{D}\!\left[\hat f^{(\ell)}\right]\tilde\rho_{c}dt\notag\\
 & +\varepsilon\sqrt{\kappa}\sum_{\ell}\mathcal{H}\left[\frac{-ie^{-i\phi}}{\kappa/2-i(\delta+\omega\ell)}\hat f^{(\ell)}\right]\tilde\rho_{c}dW(t),
\label{eq:app_good_cavity_sme}
\end{align}
where
$$
\hat H_{{\rm eff}}=\hat H_{{\rm sys}}+\sum_{\ell}\frac{\hbar \varepsilon^{2}(\delta+\omega\ell)}{(\kappa/2)^{2}\!+\!(\delta\!+\!\omega\ell)^{2}}\left(\!\hat f^{(\ell)}\hat f^{(\ell)\dagger}\!-\hat f^{(\ell)\dagger}\hat f^{(\ell)}\!\right).
$$
To enhance the signal from the QND observable $\hat f^{(0)}$ we choose
the cavity detuning $\delta=0$ and the homodyne angle $\phi=-{\pi}/{2}$.
Then, by filtering out sidebands with $\ell\neq0$ from the signal,
we obtain a homodyne current (again using Eqs.~\eqref{eq:app_photocurrent}
and \eqref{eq:app_mean_c}): 
\begin{equation}
\label{eq:app_good_cavity_Ih}
dX_{\phi}(t) \equiv I(t)dt=2\sqrt{\gamma}\langle \hat f^{(0)}\rangle _{c}dt+dW(t)
\end{equation}
with $\gamma=4\varepsilon^{2}/\kappa$ and $\varepsilon$ defined above. This gives expression for the photocurrent preceding Eq.~(6) in the main text. Discarding the sidebands from the photocurrent leads to averaging the effective SME~\eqref{eq:app_good_cavity_sme} over corresponding unobserved measurements. This results in dropping stochastic terms with $\ell\neq0$ from the equation and yields the SME~(6) in the paper. In the `good' cavity limit $\kappa\ll\omega$, the additional part in the Hamiltonian $\hat H_{\rm eff}$  is much smaller than $\hat H_{\rm sys}$ and can be neglected.

\section{Scanning Many-body Systems and the Friedel Oscillation}
\label{app:friedel}
Here we extend the scanning measurement to the many-body
case and provide the details on scanning Friedel oscillations
discussed in the main text. 

To derivation the SME describing the scan of a many-body system,
we decompose the focusing function, $\phi_{z_0}(z)=\mathcal{A}f_{z_0}(z)$ [c.f., Eq.~(1) of the main text], in terms of \emph{many-body}
eigenstates,
\begin{equation}
\sum_{i=1}^{N}f_{z_{0}}(\hat z_{i})\to\hat{f}_{z_{0}}=\sum_{\vec{\nu},\vec{\nu}'}f_{\vec{\nu}\vec{\nu}'}\ket{\vec{\nu}}\bra{\vec{\nu}'},\label{eq:f0}
\end{equation}
where $\vec{\nu}$ is the set of quantum numbers specifying
the many-body state $\ket{\vec{\nu}}$ with eigenenergy $E_{\vec{\nu}}$, 
and $f_{\vec{\nu}\vec{\nu}'}=\bra{\vec{\nu}}\sum_{i}f_{z_{0}}(\hat z_{i})\ket{\vec{\nu}'}$ are the matrix elements (Note, being a single-particle
operator, $\hat{f}_{z_{0}}$ generates only single-particle transitions).
Let us now define a set $\left\{ \Delta E_{j}\right\} $ of difference
between the eigenenergies, $\Delta E_{j}=E_{\vec{\nu}}-E_{\vec{\nu}'}$,
for all pairs of eigenstates appearing in Eq.~(\ref{eq:f0}), and define
the associated operators
\[
\quad \hat{f}^{(\Delta E_{j})}=f_{\vec{\nu}\vec{\nu}'}\ket{\vec{\nu}}\bra{\vec{\nu}'},
\]
so that $\hat{f}_{z_{0}}=\sum_{j}\hat{f}^{(\Delta E_{j})}$. Note that here we assume all $\Delta E_{j}$ being different [except
for $\Delta E_{j}=0$ corresponding to diagonal contributions of (\ref{eq:f0}){]},
as in the example of fermions in a box considered below. In situations where there are (quasi-)degenerate
energy differences $\Delta E_{j}$, like atoms in a
harmonic trap, the definition of the operators $\hat{f}^{(\Delta E_{j})}$
should include the summation over the pairs of states with (quasi)
degenerate $\Delta E_{j}$. The operators $\hat{f}^{(\Delta E_{j})}$
are generalizations of $\hat{f}^{(\ell)}$ in the
single-particle case in Appendix~\ref{app:adiabatic_elimination}, and provide a `spectral decomposition'
of $\hat{f}_{z_{0}}$: In the interaction picture
with respect to the Hamiltonian of the system, they evolve as $\hat{f}^{(\Delta E_{j})}(t)=\hat{f}^{(\Delta E_{j})}\exp(-i\Delta E_{j}t/\hbar)$. Let $\Delta E$ be a typical level spacing between physically relevant
states such that $\Delta E_{j}\geq\Delta E$. In the good cavity regime $\kappa\leq\Delta E$, these fast rotating terms with $\Delta E_{j}\neq0$ will be suppressed due to the finite time resolution $\kappa^{-1}$ of cavity, similar to the single atom case. 

We eliminate the cavity field in the same fashion as the \emph{`good cavity'} case in Appendix~\ref{app:adiabatic_elimination}. The dispersive cavity-atom coupling defines the small coeficient $\varepsilon=({\mathcal{AE}}/{\hbar })\{{\kappa}/(\kappa^2/4+\delta^{2})\}^{1/2}$. Assuming $\varepsilon\ll \kappa$ allows for eliminating the cavity in an expansion of $\varepsilon/\kappa$. Accurate to $O(\varepsilon^2)$ in the deterministic term and $O(\varepsilon)$ in the stochastic term, we arrive at the SME for the conditional density matrix of the atomic system
\begin{eqnarray}
d\tilde{\rho}_{c} & = & -\frac{i}{\hbar}[\hat{H}_{\rm eff},\tilde{\rho}_c]dt+\gamma\mathcal{D}[\hat{f}^{(0)}]\tilde{\rho}_{c}dt+\sqrt{\gamma}\mathcal{H}[\hat{f}^{(0)}]\tilde{\rho}_{c}dW(t)\nonumber \\
 &  & +\sum_{\Delta E_{j}\neq0}\gamma_{j}\mathcal{D}[\hat{f}^{(\Delta E_{j})}]\tilde{\rho}_{c}dt,\label{manybodySME}
\end{eqnarray}
where we have assumed a resonant cavity driving $\delta=0$, the homodyne angle $\phi=-\pi/2$. In Eq.~\eqref{manybodySME}, $\hat{f}^{(0)}=\hat{f}^{(\Delta E_{j}=0)}=\sum_{\vec{\nu}}f_{\vec{\nu}\vec{\nu}}\ket{\vec{\nu}}\bra{\vec{\nu}}$
 is the QND observable which measures the local density
for an arbitrary eigenstate, with a rate $\gamma=[4\mathcal{A}\mathcal{E}/(\hbar\kappa)]^{2}$. Analogous to the single-particle case
{[}c.f. Eq.~\eqref{eq:app_good_cavity_sme}{]}, the last term of
Eq.~\eqref{manybodySME} describes the suppressed dissipation channels,
with the rates $\gamma_{j}=\gamma[1+4\Delta E_{j}^2/\kappa^{2}]^{-1}$. Finally, the Hamiltonian $\hat{H}_{\rm eff}=\hat{H}_{\rm sys}+\hbar\varepsilon^2\sum_{\Delta E_j \neq 0}\Delta E_j[(\kappa/2)^2+\Delta E_j^2]^{-1}[\hat{f}^{(\Delta E_j)}\hat{f}^{(\Delta E_j)\dag}-\hat{f}^{(\Delta E_j)\dag}\hat{f}^{(\Delta E_j)}]$. The second term comes from adiabatic elimination of the cavity, and describes cavity-mediated interactions between particles. Due to the energy hierarchy $\varepsilon\ll\kappa\leq\Delta E$, this term is far smaller than $\hat{H}_{\rm sys}$ and only weakly disturbs the eigenspectrum of the system. We will neglect this tiny correction in the following discussion.

The associated expression for the homodyne current reads
\begin{equation}
I(t)=2\sqrt{\gamma}{\rm Tr}[\hat{f}^{(0)}\tilde{\rho}_{c}(t)]+\xi(t).\label{homodyneManyBody}
\end{equation}

We now apply the above analysis to a simple example of a non-interacting Fermi sea, where 
the presence of a single impurity causes the Friedel oscillation. Consider $N$ fermions in a one-dimensional box of
length $L\gg\sigma$, $-L/2\leq z\leq L/2$, with a point-like impurity
at the origin described by the potential $V_{imp}(z)=U\delta(z)$.
Assuming zero boundary conditions at $z=\pm L/2$ and taking the limit
$U\to\infty$ to simplify anlytical expressions, the single-particle
wave functions read 
\begin{equation}
\begin{split}\psi_{n}^{(o)}(z) & =\sqrt{\frac{2}{L}}\sin\left(\frac{2\pi n}{L}z\right),\\
\quad\psi_{n}^{(e)}(z) & =\sqrt{\frac{2}{L}}\sin\left(\frac{2\pi n}{L}\left|z\right|\right),
\end{split}
\end{equation}
where $n=1,\,2,$$\ldots$ for both $\psi_{n}^{(o)}(z)$ (odd parity) and $\psi_{n}^{(e)}(z)$ (even parity). The corresponding eigenenergies are
$\epsilon_{n}^{(o/e)}=[2\pi^{2}\hbar^{2}/(mL^{2})]n^{2}$. The particle density for the ground state is (we assume even $N$
for simplicity)
\begin{align}
n(z) & =\sum_{n=1}^{N/2}[\psi_{n}^{(o)}(z)^{2}+\psi_{n}^{(e)}(z)^{2}]\nonumber\\
 & =n_{F}+\frac{1}{L}\left\{ 1-\frac{\sin[2\pi(N+1)z/L]}{\sin(2\pi z/L)}\right\} ,\label{eq:n(z)general}
\end{align}
where $n_{F}=N/L$ is the average fermionic density. In the vicinity
of the impurity, $\left|z\right|\ll L/2\pi$, $n(z)$ has the form
of Friedel oscillations, 
\begin{equation}
n(z)\approx n_{F}-\frac{\sin(2k_{F}z)}{2\pi z}=n_{F}\left[1-\frac{\sin(2k_{F}z)}{2k_{F}z}\right],\label{eq:n(z) Friedel}
\end{equation}
with $k_{F}=\pi n_{F}$  the Fermi wave vector, and we omit terms $\sim L^{-1}$.
Note that for $z\sim L/2\pi$ the ``finite-size'' oscillations in
$n(z)$, Eq.~\eqref{eq:n(z)general}, have the amplitude $\sim L^{-1}$
that vanishes in the thermodynamic limit with the fixed density $n_{F}$,
in contrast to the Friedel oscillations Eq.~\eqref{eq:n(z) Friedel}.

For this case it is convenient to classify the many-body states in terms of occupations of single-particle
states and to use the language of second quantization. We introduce the destruction (thus the associated creation) operators as $\hat{b}_{n,L(R)}=\frac{1}{\sqrt{2}}\int_{-L/2}^{L/2} dz [\psi_n^{(o)*}(z)\mp\psi_n^{(e)*}(z)]\hat{\psi}(z)$ [with $\hat{\psi}(z)$ the fermi field operator], which correspond to the left(right) single-particle eigenmodes. The focusing function $f_{z_0}(\hat{z})$ has zero matrix elements between left and
right eigenmodes, $\langle m,L|f_{z_0}(\hat z)|n,R\rangle= 0$. Using these bases and for simplicity defining the single-particle quantum number $\nu\equiv\{n,L(R)\}$, Eqs.~\eqref{manybodySME} and \eqref{homodyneManyBody} can be expressed explicitly: the QND observable $\hat{f}^{(0)}$ becomes $\hat{f}^{(0)}=\sum_{\nu}f_{\nu\nu}b^\dag_{\nu}b_{\nu}$ whereas the last term of Eq.~\eqref{manybodySME} (the suppressed dissipations channels) reads $\sum_{\nu\neq \nu'}\gamma_{\nu\nu'}\mathcal{D}[\hat{b}^\dag_\nu \hat{b}_{\nu'}]$ with the corresponding rates $\gamma_{\nu\nu'}=\gamma  f_{\nu\nu'}^2[1+4(\epsilon_\nu-\epsilon_{\nu'})^2/\kappa^2]^{-1}$, where $f_{\nu\nu'}=\langle \nu |f_{z_0}(\hat z)| \nu'\rangle$ is the single-particle matrix element and $\epsilon_\nu=[2 \pi^{2}  \hbar^{2}/(mL^{2})]n^{2}$. By truncating to a suitable number of fermi orbitals, Eqs.~\eqref{manybodySME} and \eqref{homodyneManyBody} can then be integrated straightforwardly.

To resolve the Friedel oscillations in the scan, their period has
to be larger than the focusing region $\sigma$, $\pi/k_{F}>\sigma$.
This condition puts an upper bound on the density of fermions and,
therefore, on their total number, $N<L/\sigma$, which corresponds
to having not more than one fermion per length $\sigma$. The gap
to the first excited state (the level spacing) in this case can be
estimated as $\Delta E\sim\hbar^{2}/(m\sigma^{2})$, and the condition
for the non-demolition scan reads $\kappa\leq\hbar^{2}/(m\sigma^{2})$. For the simulation shown in Fig.~4b of the main text, we consider $N=16$ fermions, scanned by a microscope with resolution $\sigma=0.01L$ and cavity linewidth $\kappa=4\pi^2\hbar^2/(mL^2)$. The dimensionless measurement strength is $\gamma T=400$ with $T$ being the total scanning time. The filter integration time for post-processing is chosen as $\tau=\sigma T/L=0.01T$.


\end{document}